# Debiasing the Influence of Demographic and Appearance Cues in Social Engineering via Role-Taking: Negative Results

Tourjana Islam Supti[1*], Israa Abuelezz[1], Aya Muhanad[1], Mahmoud Barhmagi[1], Ala Yankouskaya[3], Khaled M. Khan[1], Aiman Erbad[1], Raian Ali[2]

[1]College of Engineering, Qatar University, Doha, Qatar
[2]College of Science and Engineering, Hamad Bin Khalifa University, Doha, Qatar
[3]Department of Psychology, Bournemouth University, Poole, UK
*Corresponding Author {Email: tourjana.supti@qu.edu.qa, ORCID: 0000-0002-1302-1607}

## Abstract

This study investigates the efficacy of role-taking and literacy-based interventions in reducing the influence of appearance cues, such as gender, age, ethnicity, and clothing style, on trust and risk-taking in social engineering contexts. A-4 (Group: Control, Literacy, Persuader, Persuadee) × 2 (Time: Pre, Post) mixed factorial design was implemented over two weeks with 139 participants. The control group received no material. The literacy group attended two sessions focused on how behavior can be similar regardless of appearance cues. The persuader group completed three sessions, learning how to use such cues to influence others. The persuadee group attended three sessions involving the selection, justification, and reflection on personas and scenarios. Scenarios centered on financial and rental advice. A one-week gap followed before post-intervention testing. In both pre- and post-tests, participants assessed personas combining appearance cues, offering mobile hotspots with potential risk. They rated trust and willingness to take the risk. Validated measures and scenarios were used, including word-of-mouth and issue involvement scales. It was expected that cue influence would diminish post-intervention. However, no significant within- or between-group differences emerged. Findings raise concerns about the effectiveness of debiasing efforts and call for reconsideration of approaches using literacy, role-taking, rehearsal, drama, and simulation.

## 1. Introduction

The rapid growth of technology has amplified the scope for cyberattacks, increasing security vulnerabilities across virtually every sector (Dimitrov, 2020). Human error remains the leading challenge in cybersecurity, with technological developments acting as a contributing factor (Lohani, 2019). Social engineering is one of the most persistent and successful human-centered threats compared to other cybersecurity attacks (Kaushalya et al., 2018). As a broad and manipulative technique, it exploits typical patterns of human behavior to deceive individuals into revealing sensitive information or breaching security protocols for someone else's gain (Sadiku et al., 2016). Instead of using technical vulnerabilities, hackers exploit human psychology and use emotions such as trust, fear, urgency, or obedience to authority to manipulate their victims (Siddiqi et al., 2022; Singh, 2025; Wang et al., 2021).

Social engineering attacks typically involve the use of deceptive cues delivered via emails, phone calls, or even body language to manipulate individuals into acting without due caution (Qin & Burgoon, 2007; Sadiku et al., 2016). The attackers often leverage subtle cues like looks, age, gender, or the perception of authority to gain instant trust and boost success rates. In such situations, decisions are often driven by heuristics and mental shortcuts (Gigerenzer & Gaissmaier, 2011), where trust is placed based on demographic characteristics rather than an accurate risk assessment, making social engineering one of the most successful cyber-attack strategies nowadays. This manipulation is often structured through persuasion, fabrication, and data gathering, three core dimensions that define the anatomy of social engineering attacks (Tetri & Vuorinen, 2013). Extensive research across psychology, behavioral science, and cybersecurity has demonstrated that social stereotypes, particularly those related to age, gender, and appearance, influence perceptions of trustworthiness, risk-taking, and susceptibility to manipulation (Fisk, 2016). Visual characteristics such as facial features, expressions, or style of dress are also often subconsciously linked to assumptions about an individual's intentions or reliability, making them powerful cues in social engineering contexts (Ledesma et al., 2020).

To reduce social engineering risks and enhance online safety, both technical and human-centered mitigation strategies are essential (Zulkurnain et al., 2015). One effective strategy is to raise awareness and promote greater caution in the online environment through targeted training programs that educate individuals (Smith et al., 2013). However, the effectiveness of training remains debated: some studies report significant improvements in resistance to phishing and social engineering after training, whereas others find that gains are short-lived or negligible in practice (Jampen et al., 2020; Rozema & Davis, 2025). This inconsistency highlights the need for innovative strategies that go beyond traditional literacy programs. Beyond education and digital literacy, interactive methods such as role-taking exercises can further enhance individuals' ability to recognize and resist manipulation. Role-taking offers a practical, experiential approach to improving decision-making under pressure by simulating real-world persuasion attempts.

This paper presents a role-taking strategy to increase individuals' understanding of demographic and appearance biases while strengthening their defenses against social engineering tactics. Rather than simply training participants to identify specific social engineering attacks, we aim to raise their awareness of how subtle appearance cues affect their decision-making. In an online experiment, we evaluated the effectiveness of this method in reducing vulnerability to social engineering, especially among those already conscious of security threats. Although the ethical implications of having participants assume the role of social engineers must be considered, this approach shares similarities with persuasive strategies used in marketing and advertising, making it a relevant and applicable method for combating social engineering (Fransen et al., 2015).

This paper is structured as follows: Section 2 examines the theoretical foundations underlying the study; Section 3 describes the experimental methodology; Section 4 presents the results; and Section 5 discusses the findings concerning existing literature, acknowledges the study's limitations, and outlines suggestions for future research.

# 2. Theoretical Background

## 2.1 Demographics and Appearance Stereotypes

Among the factors influencing decision-making, age is a particularly influential stereotype that affects how individuals are perceived across various contexts. For example, older individuals are often perceived as possessing positive social qualities, such as wisdom and warmth, and are often viewed as more experienced and credible (Fiske et al., 2002). Moreover, older individuals are perceived as more reliable or better suited to offer advice due to societal associations between age and knowledge (Harwood et al., 1993). Studies also indicated that emerging adults tend to be associated with energy, adaptability, and technological competence, particularly in digital or fast-changing environments (NdiBalema, 2020; X. Zhang & Zhou, 2023). There is also evidence that impulsive behavior, inexperience, and poor judgment, often associated with younger people, can contribute to lower trust toward younger individuals (Carvalho et al., 2023), especially in situations requiring careful judgment. These findings suggest that age-related stereotypes create specific expectations about how individuals behave in trust and decision-making situations. However, how strongly these assumptions influence actual behavior in applied contexts, such as cybersecurity, is still unclear.

Gender stereotypes also affect how individuals are perceived and treated in terms of trustworthiness, authority, and risk across both social and professional settings. For instance, females are commonly associated with warmth, honesty, and approachability characteristics that stem from traditional gender stereotypes positioning women as nurturing and cooperative (Ellemers, 2018; Fiske et al., 2002; Kite et al., 2008). The expectations of trust and honesty are often present in the healthcare, educational, or caregiving environments where female figures are perceived as compassionate and morally driven (Christov-Moore et al., 2014; Singer, 2023). In contrast, men are often perceived as competent, assertive, and authoritative, particularly in leadership, business, or technical environments. There is also a prevalent stereotype that men are better at mathematics than women, which can contribute to disparities in academic self-concept and performance expectations (Rossi et al., 2022). These biased perceptions of gender could create automatic cognitive shortcuts that shape responses to male and female figures in high-risk scenarios. Social engineers may exploit these stereotypes by adopting female personas to appear harmless and trustworthy, thereby reducing suspicion and increasing the likelihood of persuasion. A recent study reported that participants showed greater trust and higher risk-taking with female personas, even in social engineering scenarios with a cyber threat (Abuelezz, Barhamgi, Nhlabatsi, et al., 2024).

Stereotypes based on physical appearance also play a substantial role in how people assess trust and perceive risk. Research has shown that attractive individuals are viewed as more trustworthy, confident, and competent due to the "Halo Effect," where positive attributes are unconsciously assigned based on physical features (Batres & Shiramizu, 2023). This bias can carry over into professional and cybersecurity environments. Attire, for example, can significantly impact perceived trustworthiness; formal clothing, such as a suit and tie, often conveys authority and professionalism, while casual attire triggers perceptions of nonconformity and reduced credibility (Howlett et al., 2013). Social engineers frequently exploit such appearance-based cues, deliberately presenting themselves in ways that evoke trust or sympathy. Furthermore, research suggests that the impact of appearance cues on trust and compliance can vary by gender. For instance, females dressed casually were found to receive higher acceptance rates, whereas males were more successful when dressed formally (Abuelezz, Barhamgi, Nhlabatsi, et al., 2024).

In addition to gender, age, and physical appearance, ethnicity and background also influence how individuals develop trust in one another. When people share a common ethnicity, it gives them a sense of familiarity and belonging and more aligned feelings in their core values, making it easier to trust one another (Hansen, 2005; Wildman, 2010). This is a well-documented phenomenon explained by Social Identity Theory, which suggests that individuals categorize themselves and others into in-groups and out-groups (Tajfel & Turner, 1979). Ethnic similarity strengthens in-group identification, leading to higher trust, perceived reliability, and reduced skepticism (H. Zhang et al., 2022). In the professional world, including the digital context, we may unconsciously trust others with similar cultural behavior patterns because we assume their intentions align with ours. This can affect decision-making in hiring, financial transactions, and cybersecurity. Social engineers and cybercriminals can leverage the same dynamic to manipulate targets. For example, it was demonstrated that Arab participants were more likely to accept and trust personas with ethnic features similar to their own, illustrating how similarity in ethnicity can impact an individual's susceptibility to social engineering (Abuelezz, Barhamgi, Nhlabatsi, et al., 2024).

**2.2 Role-taking and Literacy Interventions**

Bias is difficult to eliminate because of ingrained stereotypes and unconscious influence, but it can be minimized via training, mindfulness, and structured decision-making (Brownstein, 2015; Huang & Kuo, 2015). Awareness programs and training initiatives are popular ways to reduce these biases and help people reduce susceptibility to these tactics. Serious games, scenario-based learning, and interactive immersive environments serve as educational interventions that enhance cybersecurity awareness by helping individuals recognize and resist manipulation tactics in realistic threat scenarios (Batzos et al., 2023; Shillair et al., 2022). While role-playing is an active learning technique that involves participants adopting specific roles within structured scenarios to simulate social interactions, role-taking emphasizes perspective-shifting and the cognitive process of imagining oneself in another's situation (Uzefovsky & Baron-Cohen, 2019). Both approaches allow individuals to engage with realistic challenges, reflect on their behavior, emotions, and decision-making processes, and develop practical skills. This experiential learning improves motivation and retention (Delos Santos & Fiscal, 2024; Sia et al., 2024). One of the core benefits of such methods is raising awareness of unconscious biases. These activities can help individuals understand how social cues and stereotypes may bias their decision-making (Strough et al., 2011; Tsergas, 2021), particularly relevant in contexts vulnerable to manipulation, such as cybersecurity. Participants gain insight into manipulation tactics by experiencing how trust can be shaped by appearance or group-based assumptions. These interactive methods enhance engagement and retention and build individuals' capacity to detect and resist social engineering strategies more effectively.

While some interventions place individuals in simulated scenarios to confront their biases, literacy education offers a complementary approach. It equips individuals with the knowledge and critical thinking skills needed to recognize and resist manipulation based on social biases and decision-making shortcuts. Bias literacy, in particular, focuses on helping individuals understand how stereotypes and automatic assumptions affect their judgments and provides strategies to challenge these patterns. Moreover, educational resources (e.g., structured reading materials and guides) can play a key role in this process by making complex topics more understandable (Olson, 2009). A similar approach is seen in cybersecurity literacy, highlighting how education can shift behavior by deepening awareness and critical thinking rather than relying on passive training (Ojha & Chattopadhyay, 2025). This pattern holds in health contexts as well, where literacy programs have enabled underserved populations to understand medical information better and participate more confidently in decisions about their treatment (Seo et al., 2016). Similarly, studies show that literacy-based learning can strengthen individuals' ability to think critically, analyze information, and make more informed, less biased decisions (Braunger & Lewis, 1998).

While existing literacy and role-taking interventions have shown promise in reducing implicit bias in general contexts, such as education, healthcare, or intergroup relations (Sabin, 2022; Tsergas, 2021), very few have explored their application within cybersecurity. Specifically, there is a lack of research on how role-taking can help individuals recognize and resist manipulation driven by demographic and appearance-based cues in cyber threats. This represents a significant gap in the literature, as such biases are frequently exploited by social engineers to bypass rational scrutiny.

Our study aims to address this gap by applying a role-taking approach specifically targeting how these subtle cues, such as age, gender, and appearance, impact vulnerability to manipulation in cyber threats. By increasing individuals' persuasion knowledge, we can help them develop resistance to such influences (Friestad & Wright, 1994), specifically their awareness of how manipulation works through these cues. This approach draws on inoculation theory, much like medical vaccination, which builds resistance to future manipulation by exposing individuals to weakened persuasive attempts paired with counterarguments, helping them deconstruct and recognize such strategies (Compton, 2025). Additionally, presenting conflicting information about persuasion tactics can prompt individuals to reconsider their beliefs. Exposing them to diverse viewpoints helps them acknowledge that there are multiple perspectives and that no single truth exists (Kerwer & Rosman, 2018). The mechanisms behind the use of persuasive techniques that may have been overlooked before, with the goal of enhancing individuals' understanding of persuasion and improving their ability to resist it. While previous research has shown that such strategies are effective in various domains, it often suffers from limitations like low ecological validity, reliance on self-reported measures, and immediate post-intervention assessments. Our study improves upon these by utilizing validated measures, applying stricter participant eligibility criteria to increase relatability to the study scenarios, and reassessing participants one week after the intervention to measure lasting effects. This approach not only fills a gap in the literature but also offers a more realistic, delayed assessment of how role-taking and persuasion literacy can mitigate manipulation in cybersecurity contexts.

To address this, we aim to test the following hypothesis:

**H1:** Participants in the intervention groups (role-taking and literacy-based) will demonstrate lower bias in trust and risk-related decisions across demographic contrasts (e.g., female vs. male, older vs. younger, casual vs. formal attire, Arab vs. European ethnicity) than participants in the control group.

**H0:** There will be no significant difference in the effectiveness of the role-taking and literacy-based interventions in reducing the influence of demographic and appearance-based cues.

# 3. Method

The following section describes the study design, the participants' recruitment process, face validation, measures, and statistical analyses.

### 3.1 Study Design

An online experiment using a 4×2 mixed factorial design (4 groups × 2 timepoints) was conducted across four phases via the SurveyMonkey platform (https://www.surveymonkey.com/). In Phase 0 (Eligibility Criteria), participants were screened using an eligibility survey to ensure they met the inclusion criteria. In Phase 1 (Pre-Intervention), participants' baseline reliance on appearance and demographic cues when assessing the trustworthiness of potential social engineers was measured. This phase featured scenarios in which a stranger approached the participant in a public space and offered to share their mobile internet hotspot during a moment of need. The 16 personas (please refer to section 3.3.2) were introduced in this phase to simulate realistic, trust-related decision-making situations based on key attributes, such as age, gender, ethnic similarity, and style of appearance. Those scenarios, also referred to as vignette experiments, enable researchers to present participants with controlled hypothetical situations (Steiner et al., 2017), specifically manipulating the personas' visual characteristics. This vignette-based approach is known to enhance experimental realism and is generally perceived as more engaging than traditional survey questions. For the detailed description of this phase, please refer to section 3.3.2.

In Phase 2 (Intervention), participants were randomly assigned to either the control group or one of three intervention groups (Persuader, Persuade, and Literacy) using a weighted stratified randomization method. The intervention was delivered across three sessions, except for the literacy group, which participated in two sessions. The control group, which received no intervention, was recruited separately to prevent contamination. Group allocation was based on a weighted scoring formula that prioritized gender, followed by perceived security risk and risk-taking likelihood, with final assignments manually reviewed for accuracy. Participants were unaware of the existence of multiple groups. The role-taking tasks were presented in a different context, not related to social engineering, to prevent participants from guessing what the study was about. During the intervention, they learned about stereotypes related to age, gender, and appearance through either role-taking exercises or written materials. The following subsection offers a more detailed explanation of the intervention. Phase 4 (Post-Intervention) was conducted one week after the intervention phase to reduce recall bias and capture more natural changes in participants' perceptions over time. **Figure 1** shows an overview of the study design, and details of the study design and methodology are available on the Open Science Framework (OSF) link included in the supplementary materials.

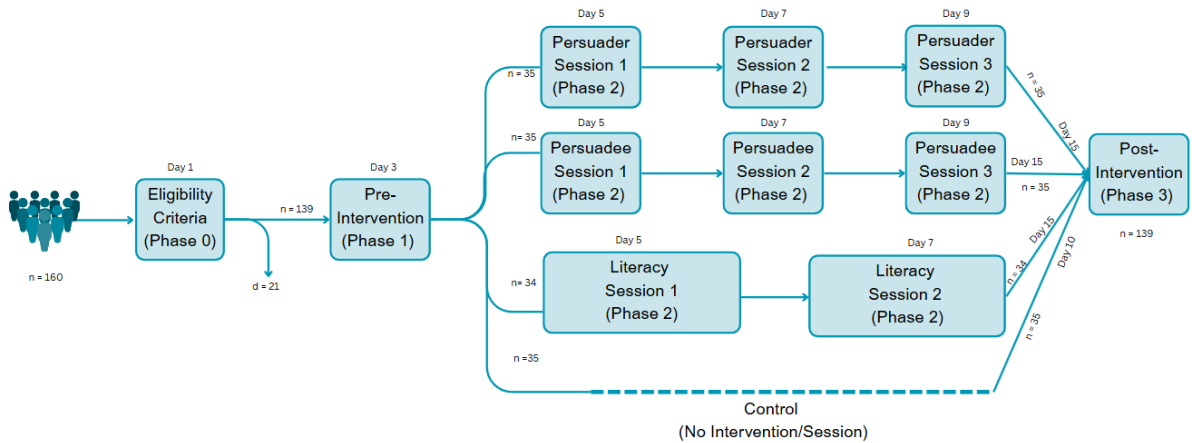

**Figure 1:** Study Design and Data Collection Timeline Overview. Note: n= Number of Participants, d= Dropout Participants

### 3.2 Participants

Participants were recruited via Prolific (https://www.prolific.com/), a platform that allows researchers to target specific groups based on demographic and behavioral criteria. To ensure cultural coherence, the study was visible only to users aged above 18 who were nationals of one of the 22 Arab countries or of Middle Eastern ethnicity with Arabic as their first language. Participants were also asked to identify with Arab cultural traits, based on Charles Harb's framework (Harb, 2016), which was explained with brief descriptions. Inclusion criteria also required participants to be fluent in English, be familiar with mobile hotspot use, perceive risk when using a stranger's hotspot, be open to receiving help from strangers, and have access to a laptop or desktop. Only those who completed all the phases received compensation. Payment was set at a "good" to "great" level to encourage engagement, with a bonus provided in the final phase as a token of appreciation.

An a priori power analysis was conducted using G*Power to ensure sufficient statistical power. The analysis assumed (4 groups × 2 timepoints), a standard alpha level of 0.05, and a small-to-moderate effect size ($\eta^2 = 0.15$), targeting 80% power ($1 - \beta = 0.80$). Results indicated a minimum sample of 128 participants was required. A 20% buffer was applied to mitigate potential attrition, yielding a final recruitment target of approximately 160 participants. A total of 139 participants were eligible, and any who failed to complete a phase were withdrawn and replaced to preserve the integrity of the experimental design. Participants were notified when each new phase was launched and were given 24 hours to complete their submissions. Ethical approval for the study was obtained from the Institutional Review Board (IRB) at the first author's institution.

### 3.3 Procedure

We distributed all study materials directly to participants via Prolific and used Prolific's chat feature to provide feedback on their submissions. Participants were required to actively engage with the intervention by completing written tasks. Informed consent was obtained prior to the screening process, and participants were reminded of their right to withdraw at any stage of the experiment. However, they were also informed that only those who completed all phases would be eligible for compensation. To maintain data integrity, attention checks were embedded throughout the study, and participants who failed multiple checks were disqualified.

### *3.3.1 Phase 0 (Eligibility Criteria)*

Participants were asked to complete eligibility questionnaires to determine if they met the inclusion criteria. Those who did not qualify based on the criteria and individuals who failed attention checks, gave inconsistent responses, or did not complete the study were excluded.

### *3.3.2 Phase 1 (Pre-Intervention)*

In this phase, participants were asked to imagine that they were sitting in a public place, such as a café, without internet access on their phone. In the scenario, a stranger notices their lack of connection and offers to share their mobile hotspot. Participants were then introduced to 16 personas (see **Figure 2**), each designed based on key attributes such as age, gender, ethnicity, and dress style (formal or casual). These personas were carefully crafted to simulate realistic, trust-related decision-making situations and were face-validated to ensure they accurately represented the intended demographic characteristics. The face validation process was conducted to ensure accurate ethnic representation and overall credibility of the personas. Individuals from diverse cultural backgrounds assessed the personas to confirm that they not only reflected European or Arab ethnic traits but were also perceived as typical in terms of likability and everyday appearance. The personas were selected based on prior research indicating that, for instance, female and older personas are generally trusted more than male and younger ones, leading to a higher willingness to accept risk in such interactions. In this study, the personas were grouped accordingly to explore how demographic features influence trust-based decisions. Participants were asked to rate how much they trusted each persona and how likely they would be to accept the hotspot offer. This allowed researchers to examine how trust and decision-making were influenced by the personas' demographic and appearance-based features. To access all scenarios and questions presented in Phase 1, please refer to the OSF link provided in the supplementary materials section.

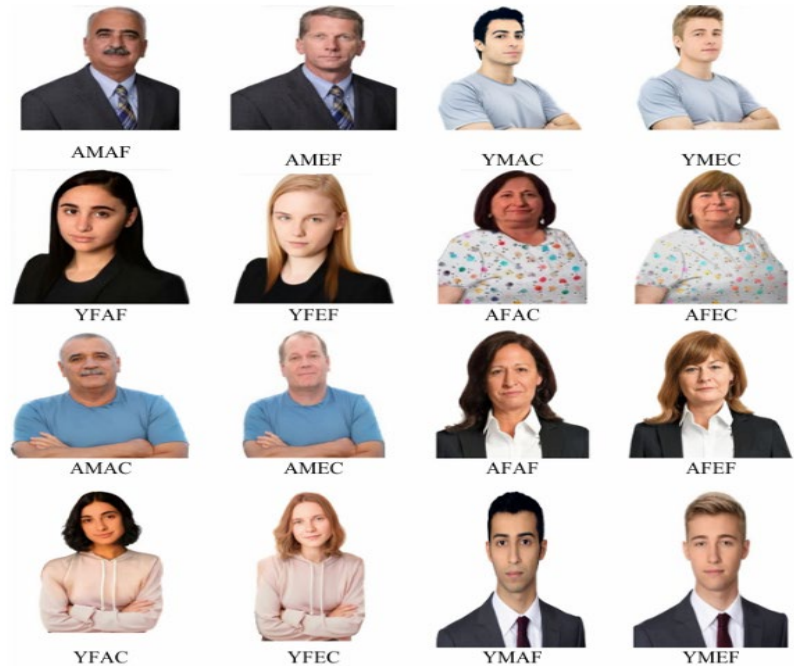

**Figure 2:** 16 Personas used in the Pre and Post Intervention Survey of the study <**A**ged | **Y**oung, **M**ale | **F**emale, **A**rab | **E**uropean, **F**ormal | **C**asual>

**Source:** Created by authors using an AI tool.

*3.3.3 Phase 2 (Intervention)*

In this phase, participants were randomly assigned to one of four groups: (1) the Role-Taking Group (Persuader), where participants acted as persuaders applying persuasive techniques; (2) the Role-Taking Group (Persuadee), where participants acted as persuadees making trust-based decisions; (3) the Literacy Group, which received educational material on the influence of demographic and appearance-based stereotypes; and (4) the Control Group, which received no intervention. The Control Group only participated in the pre- and post-intervention phases to serve as a baseline comparison. In contrast, participants in the Persuader and Persuadee groups engaged in three structured sessions during the intervention phase, while the Literacy Group completed two sessions. The intervention featured tailored role-taking scenarios and literacy-based learning modules that simulated real-world decision-making, with each session focusing on different scenarios involving demographic and appearance cues.

Scenarios were initially developed using *Storyboard*, a visual sequencing tool traditionally used in filmmaking, to map out storyline interactions and decision-making paths. To enhance visual consistency and realism, the scenarios and characters were later recreated in *Pixton*, a web-based comic design tool that allows for customizable avatars and settings. All characters were created in Pixton, with demographic attributes (age, gender, ethnicity, and style of appearance) carefully controlled to isolate one target factor per case while minimizing confounding variables. For cultural relatability, Arabic names, Rami and Sara, were chosen to reflect the Arabian participant sample. Characters were intentionally designed to appear neutral, avoiding cues related to religion, financial status, or education level.

Pilot testing with community members and students confirmed that participants consistently recognized the intended demographic attributes. In the final session, participants validated all 16-character profiles by identifying perceived attributes; responses aligned with the intended design, and no further adjustments were required. Two final financially themed scenarios, *The Financial Advisor* and *The Lettings Agent*, were selected from an initial set of six to provide consistency while simulating real-world risk. These scenarios allowed participants to evaluate how demographic similarity influenced trust and risk-taking decisions in either the persuader or persuadee role.

For the role-taking sessions, participants were randomly assigned to one of the two roles: persuader or persuadee. Across three sessions, participants engaged with realistic scenarios that simulated everyday decision-making tasks. Each session focused on different social cues (e.g., gender, age, ethnicity, and appearance) believed to influence trust formation and persuasive effectiveness. In session 1 (the financial advisor scenario), participants were introduced to a character named Rami, who was faced with selecting between two financial advisors for a high-stakes investment decision. Across three cases, the advisors differed on a single attribute: gender (male vs. female), age (older vs. younger), or ethnicity (similar vs. dissimilar to Rami). Rami's character was visually represented as a silhouette to minimize appearance-based bias. Each participant completed the task from one of two perspectives.

Participants acting as persuaders assumed the position of employees at a financial consulting agency. Their objective was to select the advisor they believed would best gain Rami's trust and successfully persuade him to invest in the agency's financial product. For example, in the gender case, participants responded to the question:

*"To gain Rami's trust and ensure he feels confident in investing, would you send Omar (male advisor) or Ameera (female advisor)?"*

Participants in the persuadee role took the perspective of Rami and selected the advisor they trusted more to act in their best financial interest. For instance, participants responded to the prompt:

*"To make the best investment decision and ensure your financial benefits are prioritized over the advisor's commission, who would you trust more: Omar or Ameera?"*

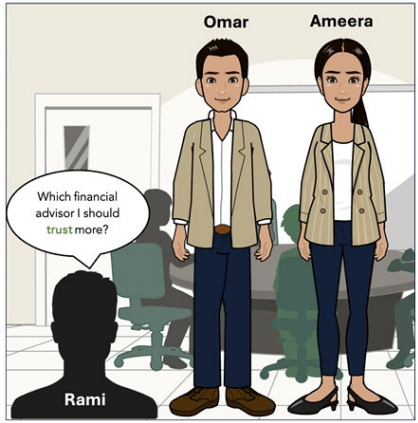 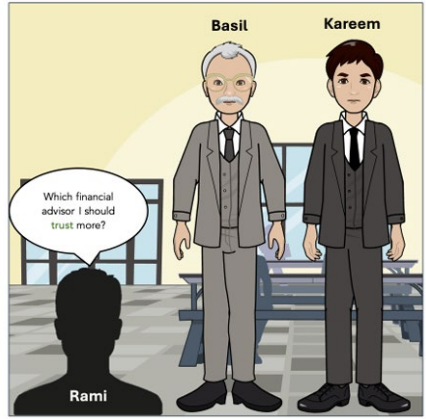 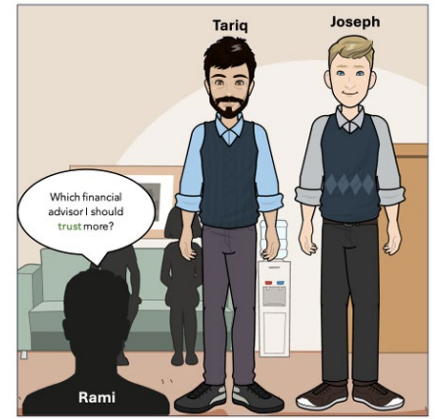

**Figure 3:** The Financial Advisor Scenarios for Role-taking

In session 2 (the lettings agent scenario), participants were introduced to a second character, Sara, who was choosing between two lettings agents to help her rent a flat. The agents varied by dress style (casual vs. formal), gender, and age. As with the previous session, Sara was depicted as a silhouette to control for potential appearance biases. Participants continued in their assigned roles.

Persuader acting as lettings agency employees, persuaders selected which agent would best earn Sara's trust and persuade her to rent one of the agency's properties. For example:

*"To gain Sara's trust and eventually agree to rent one of the available properties presented by your agency, would you send Ameer (casual style) or Majid (formal style)?"*

Persuadees assumed Sara's perspective, selecting the agent they trusted more to prioritize their housing needs and provide honest assistance. For example:

*"To make the best rental decision and ensure your needs are prioritized with honesty, who would you trust more: Ameer or Majid?"*

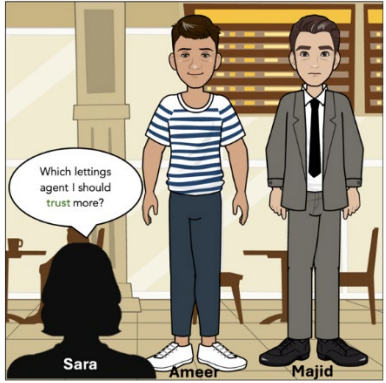 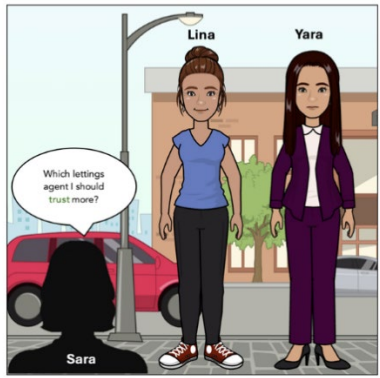 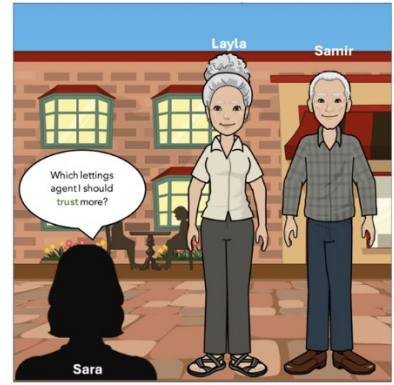

**Figure 4 :** The Lettings Agent Scenarios for Role-taking

In the final session, participants revisited both the Financial Advisor and Lettings Agent scenarios. This time, the characters Rami and Sara were fully revealed in terms of four demographic and appearance-based attributes: age, gender, ethnicity, and style of appearance. To ensure variation across scenarios and encourage more nuanced analysis, Rami and Sara were assigned distinct profiles. Specifically, Rami was presented as a young Arab male, dressed casually, and Sara was presented as an older European female, dressed formally. This design choice allowed for diversity in perceived similarity and its potential influence on trust-based judgments. In this session, participants selected one individual from a grid of 16 unique advisor/agent profiles (see Figure 5), each representing a different combination of the four attributes. These character options remained consistent across both the Financial Advisor and Lettings Agent scenarios, regardless of participant role. Participants in the persuadee role evaluated which advisor/agent they trusted most based on their own characteristics and perceived similarity. Participants in the persuader role selected which profile would most likely gain the trust of Rami or Sara, given the character's revealed attributes. This session enabled a more comprehensive analysis of how multiple demographic and appearance cues interact to shape trust formation.

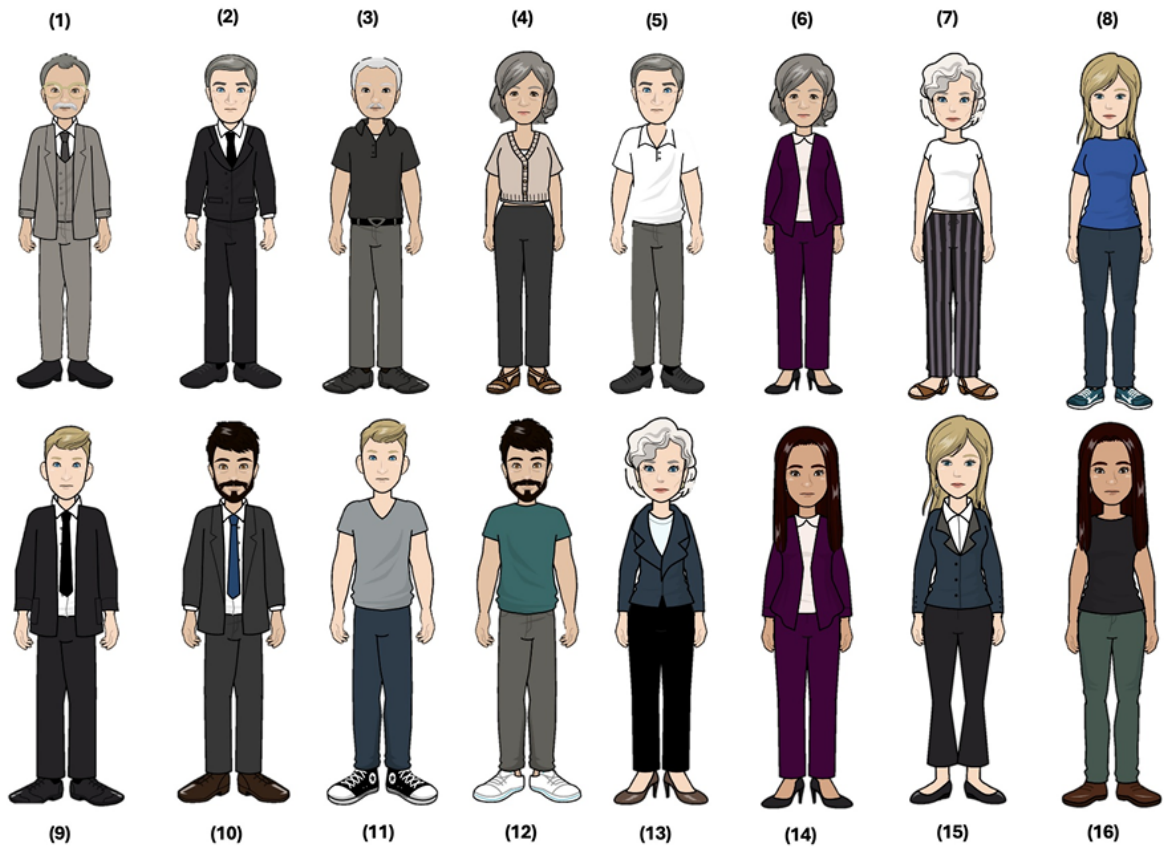

**Figure 5:** 16 Personas for Session three of Role-taking

The literacy component consisted of two structured sessions aimed at raising participants' awareness of prevalent stereotypes related to age, gender, ethnic similarity, and style of appearance, and how these perceptions can influence trust and decision-making. Drawing from peer-reviewed research, industry reports, and credible media sources, each session presented real-world examples accompanied by multiple-choice questions to encourage reflection and assess understanding. The first session focused on age- and gender-related stereotypes, exploring assumptions such as older adults being less tech-savvy, preferences in hiring based on age, and gendered perceptions of technological competence and teaching roles. The second session addressed how ethnic similarity and appearance-based judgments, like attire in professional or social contexts, can shape decisions and impressions.

To ensure a balanced and educational approach, each stereotype was countered with alternative perspectives or evidence-based challenges to the assumption. For example, while younger individuals are often perceived as more technologically adept, the session also featured research demonstrating that older adults can achieve digital literacy with appropriate support. This dual-perspective framework was applied consistently to foster critical thinking and demonstrate that stereotypes are not fixed truths. Ultimately, the goal was to help participants recognize the influence of such biases on trust-related judgments, with the intention of reducing the potential for misjudgment and promoting more equitable decision-making.

*3.3.4 Phase 3 (Post-Intervention)*

This phase replicated the procedures from Phase 1, in which participants were again asked how they would respond to the hotspot offer scenario. It was conducted one week after the completion of the intervention sessions (Phase 2, Sessions 1–3) to minimize recall bias and avoid artificially inflated responses that might result from immediate post-intervention testing. This delay allowed for a more reliable assessment of the intervention's lasting impact on participants' trust-related decision-making and perceived vulnerability, rather than capturing short-term effects or demand characteristics.

### 3.4 Measure

*3.4.1 Demographic Measure*

Participants were asked to provide demographic information, including their age and gender.

*3.4.2 Risk-taking and Trust Evaluation*

Participants' responses were assessed using a 1 to 10 Likert scale (1 = Very Unlikely, 10 = Very Likely) to evaluate risk-taking propensity and trust evaluation after each scenario in the pre-intervention phase. Risk-taking propensity measured participants' likelihood of accepting a hotspot offer from different persona groups, assessed through the question: *"On a scale from 1 to 10, how likely are you to accept the offer to use a mobile internet hotspot from a person in this group?"* Similarly, trust evaluation measured participants' perception of security risks associated with the persona offering the hotspot, using the question: *"How likely do you think a person in this group may try to compromise your data privacy and security while you are connected through their mobile internet hotspot?"* Trust scores were reversed to represent the perceived likelihood of security compromise.

To analyze bias in perceptions, responses were compared across pre- and post-intervention phases for each demographic and appearance factor. The amount of bias for risk-taking and trust was calculated as follows:

$$\text{Amount of Bias}_{(Risk\_taking\ or\ Trust),(Pre\ or\ Post)} \\ = (Risk\_taking\ or\ Trust)_{Factor\ A,(Pre\ or\ Post)} - (Risk\_taking\ or\ Trust)_{Factor\ B,(Pre\ or\ Post)}$$

Factor A represents e.g., aged, female, Arabian-looking, casually attired, and female in casual attire, while Factor B represents the corresponding counterpart e.g., young, male, European-looking, formally attired, and female in formal attire). These comparisons were conducted separately for pre- and post-intervention phases to assess changes in biases. The analysis considered differences between groups such as aged vs. young, female vs. male, aged female vs. aged male, Arabian-looking vs. European-looking, casually attired vs. formally attired, and female in casual attire vs. female in formal attire. This approach allowed for a systematic evaluation of demographic and appearance factors influencing risk-taking and trust decisions.

*3.4.3 Word of Mouth and Issue of Involvement*

The "Word of Mouth" was used to measure the likelihood that participants would discuss the influence of appearance and visual cues with others. Originally developed and widely used in marketing research to assess consumers' tendencies to share product-related opinions (Arndt, 1967; Harrison-Walker, 2001), it was adapted in this study to reflect participants' willingness to communicate their views on appearance-based influences in decision-making contexts. The scale comprised two subcomponents: Positive Word of Mouth, which captured the likelihood of participants speaking favorably or recommending conversations around the topic, and Negative Word of Mouth, which assessed the likelihood of participants expressing criticism or discouragement toward such discussions. An example item from the Positive Word of Mouth subscale was: *"What is the likelihood you will tell someone positive things about investing time and effort in understanding how demographic or appearance factors can influence decision-making and trust?"* A corresponding item from the Negative Word of Mouth subscale asked: *"What is the likelihood you will tell someone negative things about investing time and effort in understanding how demographic or appearance factors can influence decision-making and trust?"* Both items were rated on a 0–100 probability scale, where 0 indicated "no probability" and 100 indicated "certain probability."

Issue involvement was assessed using a shortened version of the Personal Involvement Inventory (PII), originally developed by Zaichkowsky (Zaichkowsky, 1985) . Commonly applied in inoculation and persuasion studies, this scale measures how personally relevant or important a particular issue is to the respondent. In the context of this study, the Issue of Involvement (IOI) was adapted to evaluate the degree to which participants considered appearance-based decision-making to be personally significant. The scale included IOI_Importance, IOI_Concern, IOI_Relevance, IOI_Meaning, IOI_Matter, and IOI_Significance. Each item was rated on a 7-point Likert scale. An example item is: *"Thinking about how a person's appearance influences my decisions is…"* rated 1 = unimportant to 7 = important. Participants' responses across all six items were summed to generate an overall issue involvement score, with higher scores indicating greater personal relevance attributed to appearance in decision-making.

All the details of the scales and questions, along with the dataset, are available in the OSF link provided in the supplementary materials.

### 3.4.4 Data analysis

All the analyses were conducted using "JASP" software version 0.19.0.0 (https://jasp-stats.org/). To assess changes in participants' risk-taking behavior and trust before and after the intervention, separate Repeated Measures ANOVAs (RM-ANOVA) are conducted for each persona factor (age, gender, ethnic similarity, and style of appearance), word of mouth, and issue of involvement across two-time points (pre- and post-intervention), resulting in fifteen independent RM-ANOVA models. The decision to analyze each persona factor independently rather than in a combined model is based on research focusing on individual persona attributes rather than their interactions. Before conducting the RM-ANOVA, normal distribution, skewness, kurtosis, and Q-Q plot are also checked. The full details regarding the skewness and kurtosis are available in Appendix A in the supplementary materials via the OSF link. Within-subject factors include time (pre- vs. post-intervention) and persona factors (age: old, young; gender: female, male; ethnic similarity: similar or different; style of appearance: formal or casual), word of mouth, and issue of involvement. The between-subjects factor is the participant role (persuader, persuadee, literacy, or control), while the dependent variables are acceptance of the hotspot offer and trust. Separate RM-ANOVA models examine how participant role, persona factors, and time influence risk-taking decisions and trust. Post-hoc analyses are conducted for significant factors with a significant threshold of α = 0.05. Effect sizes (partial η²) indicate the strength of observed effects.

## 4. Results

### 4.1 Sample characteristics

The demographic characteristics of the participants are described in Table 1.

**Table 1.** Participant Demographics

|  | Variables | Persuader (35) | Persuadee (35) | Literacy (34) | Control (35) |
|---|---|---|---|---|---|
| **Gender** | Male | 19 | 19 | 21 | 22 |
|  | Female | 16 | 16 | 13 | 13 |
| **Age** | M (SD) | 26.00(4.82) | 25.51(4.79) | 25.82(3.08) | 27.00(4.80) |
|  | Range | 19-35 | 19-35 | 19-33 | 18-35 |

### 4.2 Repeated Measure ANOVA Results

To examine whether the intervention affects participants' biases toward different persona characteristics, a series of RM-ANOVAs was conducted within subjects. This analysis tested whether there were significant changes in participants' risk-taking and trust levels from pre- to post-intervention across different roles (i.e., Persuader, Persuadee, Literacy, and Control) and persona factors (e.g., age, gender, attire, ethnicity). Additionally, Bayes Factors ($BF_{01}$) were calculated to assess the strength of evidence for the null hypothesis (H0), that is, the absence of any change due to the intervention. This approach complements traditional p-values by providing a more nuanced understanding of how likely the data are under the null hypothesis than under the alternative (H1).

The intervention did not significantly change risk-taking behavior across persona factors (Table 2). Regardless of whether participants were in the role of Persuader, Persuadee, Literacy condition, or the Control group, their level of risk-taking remained largely consistent before and after the intervention. Though statistical tests indicated no significant effects, inspection of the mean scores revealed some subtle patterns worth noting. For instance, participants in the Persuader and Persuadee roles generally exhibited slight reductions in reliance on demographic and appearance cues, such as age and gender, following the intervention. In contrast, minimal or mixed shifts were observed among those in the Literacy and Control conditions. While these changes did not reach statistical significance, they may suggest an early sensitivity to intervention content. The Bayes Factor evidence consistently supported the null hypothesis, reinforcing the interpretation that persona-based cues did not produce strong behavioral shifts in risk-taking contexts.

To further account for individual variability, the model was re-run with self-esteem, self-efficacy, and age included as covariates (See Appendix B Table II). This approach allowed us to control potential baseline differences that could influence participants' risk-taking levels independently of the intervention. However, even after adjusting for these factors, no significant interaction effects emerged (i.e., Time × Self-Efficacy, Time × Self-Esteem, Time × Age), nor was there a significant Time × Role interaction. These findings suggest that the intervention's impact remained limited regardless of participants' psychological traits or age, reinforcing the stability of appearance-based risk perceptions.

Table 2: Within-subjects Effect showing the influence of Intervention Persona RM factors on Risk-taking

| Factors | Role | Pre Intervention M(SD) | Post Intervention M(SD) | F(1,135) | p | $\eta_p^2$ | $BF_{01}$ |
|---|---|---|---|---|---|---|---|
| **Aged vs Young** | *Persuader* | 1.34(3.06) | 1.26(2.24) | 0.75 | .526 | 0.016 | 11.27 |
| | *Persuade* | 2.00(2.84) | 2.14(3.53) | | | | |
| | *Literacy* | 1.27(2.65) | 1.27(2.73) | | | | |
| | *Control* | 2.14(2.59) | 1.46(2.03) | | | | |
| **Female vs Male** | *Persuader* | 2.06(2.41) | 1.80(2.23) | 0.96 | .413 | 0.021 | 8.79 |
| | *Persuade* | 2.97(2.47) | 2.63(2.55) | | | | |
| | *Literacy* | 2.06(2.09) | 1.56(2.22) | | | | |
| | *Control* | 2.14(1.99) | 1.56(2.22) | | | | |
| **Aged Female vs Aged Male** | *Persuader* | 1.00(1.78) | 1.14(2.26) | 0.44 | .727 | 0.010 | 16.23 |
| | *Persuade* | 2.03(2.18) | 1.69(1.86) | | | | |
| | *Literacy* | 1.00(2.07) | 0.85(2.08) | | | | |
| | *Control* | 1.94(2.78) | 1.54(1.84) | | | | |
| **Arabian-looking vs European-looking** | *Persuader* | 1.60(2.30) | 1.17(2.50) | 1.03 | .381 | 0.022 | 8.23 |
| | *Persuade* | 1.71(2.15) | 1.54(2.33) | | | | |
| | *Literacy* | 0.65(1.56) | 1.06(1.89) | | | | |
| | *Control* | 0.77(2.32) | 0.49(2.59) | | | | |
| **Casually attired vs Formally attired** | *Persuader* | 0.31(2.39) | 0.26(2.17) | 0.13 | .940 | 0.003 | 19.19 |
| | *Persuade* | 1.06(2.43) | 1.09(2.53) | | | | |
| | *Literacy* | 0.00(1.76) | 0.09(1.91) | | | | |
| | *Control* | 0.26(2.34) | 0.60(2.70) | | | | |
| | *Persuader* | 0.71(1.89) | 0.91(1.93) | 0.56 | .641 | 0.012 | 13.39 |
| | *Persuade* | 1.66(1.98) | 1.23(2.01) | | | | |
| | *Literacy* | 0.41(1.46) | 0.77(1.99) | | | | |
| Female in casual attire vs Female in formal attire | Control | 0.77(2.22) | 0.69(2.27) | | | | |

**According to Wagenmakers et al. (Wagenmakers et al., 2010, 2018)* Classification scheme, $BF_{01} > 10$ provides Strong evidence for $H_0$, $3 – 10$ Moderate evidence for $H_0$, $1 – 3$ Anecdotal/weak evidence for $H_0$, 1 No evidence (data equally support $H_0$ and $H_1$), $BF_{01} < 1$ Evidence in favor of $H_1$, $BF_{01} < 1/3$ Moderate evidence for $H_1$, $BF_{01} < 1/10$ Strong evidence for $H_1$.*

Similarly, the intervention failed to significantly influence trust across the various persona factors and roles (Table 3), aligning with the pattern observed in the risk-taking analysis. In both cases, participants' behaviors remained relatively stable from pre- to post-intervention, regardless of their assignment to Persuader, Persuadee, Literacy, or Control roles. However, while risk-taking scores tended to fluctuate more inconsistently across conditions, trust-related responses exhibited slightly more apparent directional shifts in the mean values. For instance, trust in older personas increased modestly among Persuadee and Literacy roles participants. In contrast, trust in female personas declined slightly in those roles, a trend not mirrored in the risk-taking data. Moreover, both constructs showed a marginal post-intervention decline in responses toward Arabian-looking personas, although the pattern was more stable in trust. Despite these trends in the means, none of the observed differences reached statistical significance. Bayes Factor results again favored the null hypothesis, indicating moderate to strong support for the absence of an intervention effect. Thus, while trust and risk-taking appeared resistant to short-term intervention, the subtler and more consistent mean shifts in trust suggest that it may be more sensitive to demographic and visual cues than risk-taking.

To further account for individual variability, the model was re-run with self-esteem, self-efficacy, and age included as covariates (See Appendix Table II). Consistent with the primary analysis, no significant interaction effects emerged for time × role, time × self-efficacy, time × self-esteem, or time × age across persona factors, with one exception. A significant time × self-esteem interaction was observed for trust ratings toward casually versus formally attired personas ($F(1,135) = 7.57$, $p = .007$, partial $\eta^2 = .05$). This suggests that self-esteem may play a moderating role in how attire influences trust over time, albeit with a small effect size.

**Table 3: Within-subjects Effect showing the influence of Intervention Persona RM factors on Trust**

| Factors | Role | Pre Intervention M(SD) | Post Intervention M(SD) | F(1,135) | p | $\eta_p^2$ | $BF_{01}$ |
|---|---|---|---|---|---|---|---|
| **Aged vs Young** | *Persuader* | 3.20(2.93) | 3.06(2.82) | 1.14 | .337 | 0.025 | 7.30 |
| | *Persuade* | 3.60(2.83) | 4.34(3.06) | | | | |
| | *Literacy* | 2.56(3.18) | 3.06(2.46) | | | | |
| | *Control* | 3.60(2.57) | 3.20(2.11) | | | | |
| | *Persuader* | 2.09(2.05) | 1.89(2.52) | | | | |
| **Female vs Male** | *Persuade* | 3.49(1.85) | 2.89(2.18) | 0.29 | .833 | 0.006 | 18.62 |
| | *Literacy* | 2.12(2.98) | 1.44(2.49) | | | | |
| | *Control* | 2.43(2.19) | 2.20(1.86) | | | | |
| **Aged Female vs Aged Male** | *Persuader* | 1.31(1.83) | 1.20(2.15) | 0.03 | .992 | 0.001 | 24.57 |
| | *Persuade* | 1.89(2.75) | 1.86(2.32) | | | | |
| | *Literacy* | 1.18(2.34) | 1.12(2.19) | | | | |
| | *Control* | 1.43(2.79) | 1.20(2.34) | | | | |
| **Arabian-looking vs European-looking** | *Persuader* | 1.26(2.36) | 0.89(2.52) | 0.34 | .799 | 0.007 | 17.21 |
| | *Persuade* | 1.34(2.70) | 1.29(2.38) | | | | |
| | *Literacy* | 1.00(2.00) | 1.12(1.90) | | | | |
| | *Control* | 0.86(2.28) | 0.60(2.10) | | | | |
| **Casually attired vs Formally attired** | *Persuader* | 0.20(1.98) | 0.31(2.04) | 0.30 | .824 | 0.007 | 17.86 |
| | *Persuade* | 0.69(2.85) | 0.94(2.78) | | | | |
| | *Literacy* | 0.21(2.27) | 0.53(1.76) | | | | |
| | *Control* | 0.03(2.33) | 0.69(2.68) | | | | |
| **Female in casual attire vs Female in formal attire** | *Persuader* | 1.06(2.18) | 1.14(2.45) | 0.36 | .785 | 0.008 | 16.49 |
| | *Persuade* | 1.40(1.99) | 1.29(2.38) | | | | |
| | *Literacy* | 0.82(1.68) | 0.97(2.18) | | | | |
| | *Control* | 1.03(2.63) | 0.51(2.06) | | | | |

The intervention had a limited effect on participants' word-of-mouth (WOM) behaviors (Table 4). Mean scores for positive and negative WOM remained relatively stable across roles, and the differences from pre- to post-intervention were not statistically significant. There was a slight increase in positive WOM among participants in the Persuader and Literacy roles, suggesting a mild boost in willingness to endorse the persona post-intervention. Conversely, the Persuadee group showed a slight decline in positive WOM. Negative WOM showed minor, inconsistent changes, with no clear directionality across roles, suggesting that the intervention did not significantly impact participants' critical evaluations or complaints.

Unlike the results observed for trust, risk-taking, and word-of-mouth, the analysis revealed a significant effect of the intervention on the Issue of Involvement (Table 4). This indicates that participants experienced a meaningful change in how personally engaged or connected they felt with the personas or the underlying issues from pre- to post-intervention. Specifically, their involvement levels decreased after the intervention, suggesting that taking on the role of persuading may have led to psychological distancing or reduced emotional investment. This finding contrasts with the stability observed in trust and risk-taking, where no significant changes emerged across any persona factor or role. However, despite this significant overall effect, post hoc comparisons did not reveal significant differences between specific roles, implying that a particular group did not drive the observed shift in involvement but may reflect a general trend across conditions. This pattern diverges from the null effects of behavioral measures like trust and risk-taking.

When the model was re-run with self-esteem, self-efficacy, and age as covariates, no significant interaction effects were found for Positive Word of Mouth and Negative Word of Mouth across time or roles (See Appendix B Table III). Similarly, Time × Self-efficacy, Time × Self-esteem, and Time × Age interactions were not significant for any of the three constructs. However, a significant Time × Role interaction was observed for Issue of Involvement, and this effect remained consistent regardless of whether covariates were included. This robustness suggests that the intervention's influence on participants' engagement with the issue was not dependent on individual differences in self-esteem, self-efficacy, or age.

**Table 4: Within-subjects Effect showing the influence of Intervention Persona RM factors on Word of Mouth and Issue of Involvement**

| **Factors** | **Role** | **Pre Intervention M(SD)** | **Post Intervention M(SD)** | **F(1,135)** | **p** | **$\eta_p^2$** | **$BF_{01}$** |
|---|---|---|---|---|---|---|---|
| **Positive Word of Mouth** | *Persuader* | 66.00(19.13) | 71.14(18.11) | 0.74 | .531 | 0.027 | 12.39 |
| | *Persuade* | 63.43(27.00) | 62.57(23.68) | | | | |
| | *Literacy* | 63.24(26.94) | 69.71(21.67) | | | | |
| | *Control* | 62.57(23.81) | 66.29(21.02) | | | | |
| **Negative Word of Mouth** | *Persuader* | 22.29(26.58) | 24.86(25.13) | 0.53 | .660 | 0.012 | 14.13 |
| | *Persuade* | 23.71(22.76) | 21.14(19.52) | | | | |
| | *Literacy* | 22.94(25.17) | 27.06(26.80) | | | | |
| | *Control* | 28.57(22.77) | 28.57(20.60) | | | | |
| **Issue of Involvement Total** | *Persuader* | 31.57(6.12) | 30.37(5.92) | 3.18 | .026 | 0.066 | 1.00 |
| | *Persuade* | 28.86(8.29) | 31.17(6.89) | | | | |
| | *Literacy* | 31.71(6.86) | 29.18(9.05) | | | | |
| | *Control* | 30.37(6.09) | 28.46(7.37) | | | | |

## 5. Discussion

The present study aimed to test whether role-taking and literacy interventions could reduce the influence of demographic and appearance-based cues on trust and risk-taking behaviors in social engineering, directly addressing our hypothesis (H1). Contrary to H1, the results did not reveal significant changes in participants' behaviors across pre- and post-intervention conditions. Specifically, there were no substantial shifts in participants' trust or risk-taking responses to persona characteristics such as age, gender, ethnicity, or attire. These findings support the null hypothesis (H0), indicating that the interventions did not significantly alter reliance on demographic and appearance cues. While negative, these results provide important insights into the challenges of modifying deeply ingrained cognitive biases.

Previous research has shown that demographic cues, including age, gender, and ethnicity, significantly influence individuals' perceptions of trustworthiness and risk, particularly in social engineering contexts (Abuelezz, Barhamgi, El Houki, et al., 2024). Although interventions such as literacy, education, mindfulness, and structured decision-making strategies can reduce some biases (Brownstein, 2015; Carnes et al., 2012), our results demonstrate that these approaches did not significantly affect participants' trust or risk-taking in the current context. This lack of change may reflect the enduring nature of cognitive shortcuts that operate automatically and unconsciously. Rather than disappearing with brief interventions, biases linked to demographic characteristics may persist because they serve a functional role in decision-making, acting as heuristics that help individuals navigate complex or uncertain environments. A study further supports this notion, showing that individuals often comply with risky requests not due to lack of awareness, but because persuasion tactics exploit heuristics, biases, and social influences that override rational risk assessments (Muhanad, Abuelezz, et al., 2025). While bias can be managed to some extent, it is an inherent aspect of human cognition and thus cannot be entirely avoided.

Theoretical perspectives further explain the observed outcomes. Theory of Planned Behavior by Azjen, which suggests that intentions do not always translate into actions (Ajzen, 1985). Although participants may have intended to be more cautious, their levels of trust and risk-taking did not significantly decrease post-intervention, reflecting the common gap between awareness and behavior. This aligns with the Cognitive Load Theory by Sweller (Sweller, 1994), which states individuals are under cognitive load, for example, during complex or ambiguous social engineering simulations in our context, they tend to rely more heavily on mental shortcuts and stereotypes (Montañez et al., 2020). The concept of *Status Quo Bias*, introduced by Samuelson and Zeckhauser (Samuelson & Zeckhauser, 1988) may further explain the observed resistance to behavioral change. This bias refers to individuals' tendency to favor existing patterns of behavior or decision-making, even when alternatives could yield better outcomes. Such inertia may stem from loss aversion, cognitive effort required to adopt new behaviors, and the psychological comfort associated with familiar routines.

Importantly, participants' responses to persuasion cues remained unchanged across all three interventions. Despite using both educational and immersive methods, neither effectively reduced the influence of cognitive biases and heuristics on decision-making. This consistency was observed across all demographics and appearance factors and measured outcomes, trust, and risk-taking, further confirming the stability of our design. Although we did not observe statistically significant behavioral changes between pre- and post-intervention conditions, these results offer valuable insights into the complexity of behavioral adaptation, particularly in the context of deeply rooted cognitive and social biases. It is unlikely that the lack of significant effects in our study was due to insufficient statistical power. Previous research has shown that comparable training interventions generally yield medium effect sizes, even when conducted with relatively small sample sizes (Fritz et al., 2012; Gignac & Szodorai, 2016). The repeated measures design and the inclusion of Bayes Factors provide strong evidence for H0, suggesting that the lack of intervention effects is a robust finding rather than a statistical anomaly (van den Bergh et al., 2023). This emphasizes that short-term interventions may raise awareness but may not be sufficient to produce measurable behavioral change in applied cybersecurity settings. Interestingly, one interpretation of our findings is that increased awareness of these security contexts may lead individuals to believe they are better equipped to recognize and resist social engineering attempts. However, this perceived immunity might paradoxically increase susceptibility. In other words, because participants believe they are aware of the influences of those demographics and appearance cues, they may assume they are no longer influenced by them, leading to overconfidence and increased risk-taking. This is consistent with research showing that individuals with higher cybersecurity awareness or stronger security attitudes sometimes take more risks, believing that their knowledge will shield them from actual harm (Muhanad, Supti, et al., 2025).

It is important to note that the intervention's relatively short duration may have limited its ability to shift deeply ingrained behavioral patterns. While the over two-week format offered experimental control, it may not have provided the sustained exposure necessary to modify implicit biases meaningfully. Prior research suggests that such biases resist change and often require long-term engagement, repeated exposure to counter-stereotypical cues, and immersive experiences (Devine et al., 2012). This aligns with findings that short-term interventions may raise awareness without immediately producing behavioral change. Supporting this, systematic reviews have concluded that current data do not reliably identify effective interventions to reduce implicit biases, with many showing no impact or, in some cases, even reinforcing bias. As such, caution is advised when implementing bias-reduction programs, and further research is needed to evaluate their long-term efficacy (FitzGerald et al., 2019). Emerging research suggests that integrating change management principles may offer more sustainable bias-reduction strategies by targeting both individual behavior and the surrounding social environment (Nguyen-Phuong-Mai, 2021). Thus, our results do not indicate intervention failure but highlight the complexity of human cognition and the entrenched nature of appearance-based judgments.

In line with qualitative research emphasizing participant-centered evaluation (Braun & Clarke, 2006), we collected and thematically analyzed open-ended written feedback from individuals who engaged with the intervention as both persuaders and persuadees. Many participants expressed that the study was "eye-opening," "fun," and "engaging," with one noting, "I learned quite a lot about myself. Great experience - I give it 10 out of 10." This aligns with previous findings highlighting how interactive, scenario-based interventions can provoke reflection and increase awareness of implicit biases (FitzGerald et al., 2019). Participants appreciated the inclusion of breaks, reminders, and attention checks, describing the experience as "very well structured" and "thoughtfully designed." Others suggested improvements such as adding diverse images or reducing repetitiveness, echoing calls in intervention design literature for greater ecological validity and participant engagement (Fowlkes, 2005; Want, 2014). Notably, participants from the literacy intervention group recommended more interactive elements such as "decision-making games," which supports findings by Green & Bavelier regarding the cognitive benefits of gamified learning (Green & Bavelier, 2012). One participant shared, "I always think about it in my daily life and reassess my decisions," suggesting that the intervention led to lasting cognitive changes in how individuals approach their decision-making. We can frame these insights under themes such as *"Cognitive Impact, Engagement, and Participant Empowerment"*, which resonate with broader trends in intervention-based research. Even if biased behavior was not immediately reduced, these insights suggest that the intervention enhanced awareness and successfully opened avenues for future improvements in its design.

A key limitation of this study is that participants had to imagine scenarios rather than engage in real-life experiences, which may have affected the authenticity of their reactions. To improve ecological validity, we excluded participants unfamiliar with mobile hotspot usage, unwilling to accept help from strangers, or who had formal marketing training. Additionally, we excluded participants who did not hold stereotypes toward demographic groups, ensuring that the study focused on those more likely to react to demographic and appearance cues. The structure of the study, involving multiple phases, raised concerns about participant attention. We addressed this by including comprehensive attention checks to identify and exclude inattentive participants. Another limitation involved the potential for participants to recall information from the intervention too vividly, leading to an "immediacy effect." To minimize this, we remeasured vulnerability one week after the intervention, striking a balance between immediate recall and dropout risk. Social desirability bias and demand characteristics may have influenced participants to provide responses they thought were expected. We attempted to reduce this by embedding the intervention in a related context but distinct from the assessed scenarios, and by including filler questions to mask the study's purpose. However, this introduced a limitation, as participants may have struggled to apply their learned knowledge to the target context fully. Finally, while comic-based simulations (via Pixton) added structure and consistency, the static format may have constrained participants' emotional engagement. Prior research by Abawajy similarly emphasizes that the format and level of engagement in awareness delivery are critical for influencing behavior, suggesting that more interactive or multi-modal approaches (e.g., game- or video-based training) may be necessary to challenge deep-seated biases effectively (Abawajy, 2014). Future studies could explore more dynamic and interactive platforms, such as gamified modules or virtual reality simulations, to better mimic real-time decision-making under uncertainty.

Our study employed a complex and interactive intervention design to debias participants and reduce risk-taking tendencies. The design, including role-taking and scenario-based elements, was crafted to engage individuals in self-reflection and challenge their biases. While the results did not show significant debiasing effects, participant feedback suggests that role-taking and forum theater techniques could be effective in organizational settings (Jacob et al., 2019), where they might encourage greater awareness and discussion of biases. However, these methods alone may not achieve the desired debiasing outcomes, as biases are often deeply ingrained in individuals. This suggests that alternative intervention approaches may be necessary to address these biases more effectively, potentially incorporating longer-term strategies or more immersive techniques to foster deeper, lasting changes in behavior.

## 6. Conclusion

In conclusion, this study explored the effectiveness of an interactive intervention designed to debias individuals and reduce their risk-taking tendencies. While the results did not show significant debiasing effects, participant feedback highlighted the potential of role-taking and scenario-based methods in raising awareness of biases, particularly within organizational contexts. The complex and engaging design of the intervention was generally well received, with many participants noting its value in provoking self-reflection and offering an enjoyable experience. However, the findings suggest that more ingrained biases may require alternative approaches for lasting change. Future research should explore alternative intervention designs incorporating long-term, immersive experiences and further refine methods for effectively addressing deep-rooted biases. Overall, this study contributes to the growing body of literature on bias reduction and risk-taking behavior, offering insights into the limitations of current interventions and the need for continuous innovation in this field.

## Author Contribution

The study was conceptualized by RA, MB, KK and AE. The study material was then designed by IA supervised by MB, RA, KK. Study material preparation and data curation was done by IA and TIS. The statistical analysis was carried out by TIS and AM and validated and revised by AY who also contributed to conceptualization, psychology background and guided the analysis methodology. The manuscript, including preparing the theoretical underpinning, was written by TIS. The manuscript was reviewed and revised by MB, RA, TIS, IA, AY, KK and AE.

## Acknowledgment

This research was made possible by an NPRP-14-Cluster Grant # NPRP 14C-0916-210015 from the Qatar National Research Fund (a member of the Qatar Foundation). The findings herein reflect the work and are solely the authors' responsibility. Open Access Fund has been provided by Qatar National Library.

## Conflict of Interest

The authors report that there are no competing interests to declare.

## Supplementary Materials

The dataset associated with this work is uploaded alongside the appendix and other supplementary material for this article at: https://osf.io/ga95d/?view_only=66fabb6facf64c549e2c609617bb6e22

Kerwer, M., & Rosman, T. (2018). Mechanisms of Epistemic Change-Under Which Circumstances Does Diverging Information Support Epistemic Development? *Frontiers Frontiers in Psychology*, *9*. https://doi.org/10.3389/fpsyg.2018.02278

Kite, M. E., Deaux, K., & Haines, E. L. (2008). Gender stereotypes. In *Psychology of women: A handbook of issues and theories, 2nd ed* (pp. 205–236). Praeger Publishers/Greenwood Publishing Group.

Ledesma, A., Kraut, D., Quezada, R., Zayas, A., & Cano-Ruiz, M. E. (2020). Physical Appearance and Its Effect on Trust. *Journal of Emerging Investigators*. https://doi.org/10.59720/20-079

Lohani, S. (2019). Social Engineering: Hacking into Humans. *International Journal of Advanced Studies of Scientific Research*, *4*.

Montañez, R., Golob, E., & Xu, S. (2020). Human Cognition Through the Lens of Social Engineering Cyberattacks. *Frontiers in Psychology*, *11*. https://doi.org/10.3389/fpsyg.2020.01755

Muhanad, A., Abuelezz, I., Khan, K., & Ali, R. (2025). On How Cialdini's Persuasion Principles Influence Individuals in the Context of Social Engineering: A Qualitative Study. In M. Barhamgi, H. Wang, & X. Wang (Eds.), *Web Information Systems Engineering – WISE 2024* (pp. 373–388). Springer Nature. https://doi.org/10.1007/978-981-96-0570-5_27

Muhanad, A., Supti, T. I., Abuelezz, I., Yankouskaya, A., Khan, K. M., Barhamgi, M., Nhlabatsi, A., & Ali, R. (2025). Does security attitude really predict susceptibility to persuasion tactics in social engineering attempts? *Information and Computer Security*. https://doi.org/10.1108/ICS-11-2024-0280

NdiBalema, P. (2020). Unlocking the Potential of ICT for Transformative Learning among Youth: A Path to 21st Century Competencies. *Journal of Educational Technology and Online Learning*, *3*(3), 245–271. https://doi.org/10.31681/jetol.777647

Nguyen-Phuong-Mai, M. (2021). What Bias Management Can Learn From Change Management? Utilizing Change Framework to Review and Explore Bias Strategies. *Frontiers in Psychology*, *12*. https://doi.org/10.3389/fpsyg.2021.644145

Ojha, E., & Chattopadhyay, D. K. N. (2025). *Cybersecurity Education and Awareness: A Framework-Based Approach For Digital Literacy*. *11*(6). https://doi.org/10.36713/epra22511

Olson, D. R. (2009). Education and literacy. *Journal for the Study of Education and Development*, *32*(2), 141–151. https://doi.org/10.1174/021037009788001824
28